\documentclass[12pt, a4paper]{article}
\pdfoutput=1

\usepackage[height=22cm,width=16cm, centering]{geometry}
\usepackage{setspace}

\usepackage[utf8]{inputenc}

\usepackage{amsmath, comment}
\usepackage{amssymb}
\usepackage[dvipdfmx]{graphicx}
\usepackage{color}
\usepackage[sort&compress, numbers, merge]{natbib}

\definecolor{purple}{rgb}{0.5 ,0, 0.7}
\definecolor{bluegreen}{rgb}{0, 0.45, 0.35}
\definecolor{sakura}{rgb}{1 ,0.52, 0.74}
\definecolor{wakakusa}{rgb}{0.45 ,0.74, 0}
\definecolor{brown}{rgb}{0.48 ,0.23, 0}
\definecolor{skyblue}{rgb}{0.21 ,0.7, 1.}
\definecolor{purplegray}{rgb}{0.35,0.35,0.73}
\usepackage{hyperref}
\hypersetup{colorlinks=true, linkcolor=bluegreen, citecolor=purple, urlcolor=blue}

\makeatletter
\let\MYcaption\@makecaption
\makeatother

\usepackage{subcaption}
\makeatletter
\let\@makecaption\MYcaption
\makeatother

\def\ns{n_{\text{s}}}

\allowdisplaybreaks

\begin{document}

\begin{titlepage}
\begin{center}
\leavevmode \\
\vspace{ 0cm}

\hfill {\small CTPU-PTC-19-29}\\

\noindent
\vskip 2 cm
 {\Large Trans-Planckian censorship and single-field inflaton potential}

\vglue .6in

{
Kenji Kadota, Chang Sub Shin, Takahiro Terada, Gansukh Tumurtushaa
}

\vglue.3in

\textit{  Center for Theoretical Physics of the Universe, \\ Institute for Basic Science (IBS),
  Daejeon, 34126, Korea\\
}

\end{center}

\vglue 0.7in

\begin{abstract}
It was recently proposed that a field theory cannot be consistent with quantum gravity if  it allows a mode shorter than the Planck length to exit the Hubble horizon. This is called the Trans-Planckian Censorship Conjecture (TCC).
We discuss the implications of the TCC on the possible shape of the inflaton potential in single-field slow-roll inflation. 
We point out that (1) there is generically an initial condition in which the total e-folding number $N_\text{total}$ is doubled or more compared to the e-folds necessary for the cosmic microwave background fluctuations, and (2) a sizable negative running of spectral index is generically expected to make $N_\text{total}$ small. 
In concrete setups, we find a stringent constraint on the inflationary energy scale, $V_\text{inf}^{1/4} < \mathcal{O}(10) \, \text{TeV}$ with $r < \mathcal{O}(10^{-50})$, and the running parameter is bounded above as $\alpha_\text{s} \lesssim - 4 \times 10^{-3}$.
\end{abstract}
\end{titlepage}

\newpage

\section{Introduction} \label{sec:intro}

Cosmic inflation solves the horizon, flatness, and monopole problems of the Big Bang cosmology by postulating a quasi-exponential (i.e.~quasi-de Sitter (dS)) accelerated expansion period.  The quantum fluctuations in the quasi-de Sitter spacetime are squeezed after exiting the horizon, and they become classic fluctuations.  In this way, inflation also provides the seeds of large scale structure of the universe.  The predictions of  vanilla single-field slow-roll inflation such as adiabatic, Gaussian, almost scale-invariant perturbations are so far consistent with the precise observations of the cosmic microwave background (CMB)~\cite{Akrami:2018odb}.  We do not yet know, however, the detailed properties such as the inflationary energy scale and the canonical inflaton field range which is traversed during inflation.

Compared to the observational status of inflation, its theoretical ground remains less established.  For example, it is known to be difficult to construct a dS spacetime in Superstring Theory.  In particular, a conjecture to prohibit dS was recently proposed~\cite{Obied:2018sgi} and subsequently refined~\cite{Andriot:2018wzk, Dvali:2018fqu, Andriot:2018mav, Garg:2018reu, Ooguri:2018wrx, Rudelius:2019cfh}.   It requires
\begin{align}
|\nabla V|  \geq  c V  \quad\quad \textrm{or} \quad\quad   \min \left( \nabla_i \nabla_j V \right) \leq  - c' V,  \label{dS}
\end{align}
where $c, \, c' >0$ are some universal constants of order unity, and we use the reduced Planck unit. 
  The conjecture is one of the Swampland conjectures~\cite{Vafa:2005ui, Ooguri:2006in}: a field theory which cannot be consistently coupled to a quantum theory of gravity is said to belong to the Swampland (rather than the Landscape of Superstring Theory).  It was soon pointed out that the Swampland dS conjecture is in tension with the inflationary interpretation of the CMB data depending on the  numerical factors in the conjecture~\cite{Agrawal:2018own, Achucarro:2018vey, Garg:2018reu, Kinney:2018nny, Brahma:2018hrd, Das:2018hqy, Fukuda:2018haz,Ashoorioon:2018sqb}.

There is also the Swampland distance conjecture~\cite{Ooguri:2006in}.  The statement is as follows: when the field moves longer than the Planck distance, $\Delta \phi \gg 1$, there appear towers of light particles whose masses scale as $m \sim e^{- c''' \Delta \phi}$ where $c'''$ is an $\mathcal{O}(1)$ constant, which would modify the effective field theory. Thus, it effectively requires
\begin{align}
 \Delta \phi < c'' , \label{distance}
\end{align}
where $\Delta \phi$ is the field distance, and $c''$ is another $\mathcal{O}(1)$ constant.
 Applied to the inflaton, this can forbid large-field models of inflation.

Recently, Bedroya and Vafa proposed the Trans-Planckian Censorship Conjecture (TCC) \cite{Bedroya:2019snp, Bedroya:2019tba}:  If a theory has a mode which has initially a sub-Planckian wavelength and  afterwards exits the Hubble horizon, the theory belongs to the Swampland.  This is a change of viewpoint on the trans-Planckian problem of inflation~\cite{Martin:2000xs, Brandenberger:2000wr, Brandenberger:2012aj, Kaloper:2002cs, Easther:2002xe,Ashoorioon:2004vm}, which itself has been known for a while.  The TCC may seem to lack theoretical supports at first sight, but it has nontrivial relations, for instance, with other Swampland conjectures, energy conditions, and the scrambling time \cite{Bedroya:2019snp}.  The TCC puts an upper bound on  the e-folding number for a given inflationary energy, thus forbidding a complete dS, in a similar spirit with the dS conjecture (see also Refs.~\cite{Dvali:2013eja, Dvali:2014gua, Dvali:2017eba, Dvali:2018fqu, Dvali:2018jhn}). 
Another consequence of the TCC is an upper bound on the inflationary energy scale $V^{1/4} < 6.9 \times 10^8$~GeV and the flatness of the potential $\epsilon < 1.3 \times 10^{-32}$~\cite{Bedroya:2019tba}.  Since typical (polynomial) large-field models predict a much larger energy scale along with a large tensor-to-scalar ratio $r = 16 \epsilon$, the TCC seems to imply  small-field inflation, which is consistent with the distance conjecture.  See Refs.~\cite{Cai:2019hge, Tenkanen:2019wsd, Das:2019hto, Mizuno:2019bxy, Brahma:2019unn, Dhuria:2019oyf, Torabian:2019zms, Cai:2019igo,Schmitz:2019uti} for other discussions on the TCC.

In this paper, we discuss what kinds of small-field inflation potentials are consistent both with the TCC and the CMB data. We do not assume the de Sitter conjecture because it seems too strong a constraint in the field space except for the asymptotic region (We come back to this point in the final section).  Differently from the other Swampland conjectures, the TCC gives indirect constraints on the parameters in the Lagrangian through a direct constraint on the e-folding number for any physically possible initial conditions, so we will explore further consequences of the conjecture.  
After a brief review on the TCC in Sec~\ref{sec:review}, we discuss the viability of typical inflation models particularly focusing on a model whose potential has a limited flat region and monotonically increases in the region relevant for inflation.   Two main conclusions drawn from Sec.~\ref{sec:main} is that we expect a sizable negative running spectral index and that the maximally possible total e-folding number is typically twice or more compared with that required to explain the CMB fluctuations.  A summary and discussions are given in Sec.~\ref{sec:discussion}.  
  In Appendix~\ref{sec:rolling-up}, we consider an initial condition not discussed in the main text. Appendix~\ref{sec:high-order-breaking} generalizes the analyses in Sec.~\ref{sec:alpha-attractor}.  The maximum e-folding number for thermal inflation and its relation to the TCC are discussed in Appendix~\ref{sec:thermal}.

\section{Trans-Planckian Censorship Conjecture} \label{sec:review}
Here, we review the TCC. 
Quantitatively, it is  given by
\begin{align}
N_\text{total} < \log \left( \frac{M_\text{P}}{H_f} \right),  \label{TCC}
\end{align}
where $N_\text{total}= \ln (a_f / a_i)$ is the total e-folding number in a given period, $H=\dot{a}/a$ is the Hubble variable, $M_\text{P} = 1$ is the reduced Planck mass, and the subscripts $i$ and $f$ denote the initial and final time.  For applications to inflationary cosmology, we often regard $N_\text{total}$ as the total e-folding number of inflation and  $H_f$ as the Hubble parameter during inflation, or more precisely that at the end of inflation $H_\text{inf}$.  

The above inequality is an upper bound on $N_\text{total}$, but there is a lower bound on $N_\text{total}$ which requires successful inflation.  Namely, there is the condition that the mode whose length is the current horizon size must be within the Hubble horizon at some time during inflation:
\begin{align}
N_\text{total} > N_\text{hor},    \label{horizon}
\end{align}
with
\begin{align}
H_0^{-1} =  e^{N_\text{hor}} \left( \frac{90 H_\text{inf}^2 }{g_* (T_\text{R}) \pi^2 T_\text{R}^4 } \right)^{\frac{1}{3(1+w)}} \left( \frac{g_{*,s}(T_\text{R})}{g_{*,s}(T_0)} \right)^{\frac{1}{3}} \left(\frac{T_\text{R}}{T_0}\right) H_\text{inf}^{-1},
\end{align}
where for simplicity we neglect the time dependence of  $H_\text{inf}$; $g_*(T)$ and $g_{*,s}(T)$ are the effective relativistic degrees of freedom for energy density and entropy density, respectively; $w$ is the equation-of-state parameter during reheating, which we take as a constant, and $T_\text{R}$ and $T_0$ are the reheating temperature and the current temperature.  

 The condition on $V$ that there is some window for $N_\text{total}$ opening between these inequalities is 
\begin{align}
   \frac{V}{3 H_0}  \left(  \frac{g_{*}(T_\text{R}) \pi^2 T_\text{R}^4}{30 V} \right)^{\frac{1}{3+3w}} \left(  \frac{g_{*,s}(T_0)}{g_{*,s}(T_\text{R})} \right)^{\frac{1}{3}} \frac{T_0}{T_R} \leq 1.
\end{align}
In particular, for the matter-dominated era $w=0$ before reheating (realized by coherent oscillation of inflaton), 
\begin{align}
V < (6.9 \times 10^8 \, \text{GeV})^4 \left( \frac{T_\text{R}}{2.8 \times 10^8 \,\text{GeV}} \right)^{-\frac{1}{2}}  \left( \frac{g_{*} (T_R)}{106.75} \right)^{-\frac{1}{2}}  \left( \frac{g_{*,s} (T_R)}{106.75} \right)^{\frac{1}{2}}   ,  \label{V_upper-bound}
\end{align}
where we have used $H_0 = 67.27 \text{km/s/Mpc}$~\cite{Aghanim:2018eyx} and $T_0 = 2.725 \text{K}$~\cite{2009ApJ...707..916F}.
These bounds translate to the upper bound on the slow-roll parameter $\epsilon = (1/2) (V'/V)^2$, 
\begin{align}
\epsilon < 1.3 \times 10^{-32} \left( \frac{T_\text{R}}{2.8 \times 10^8 \,\text{GeV}} \right)^{-\frac{1}{2}}  \left( \frac{g_{*} (T_R)}{106.75} \right)^{-\frac{1}{2}}  \left( \frac{g_{*,s} (T_R)}{106.75} \right)^{\frac{1}{2}} ,
\end{align}
where this representative value corresponds to the upper bound on the tensor-to-scalar ratio $r = 16 \epsilon < 2.1 \times 10^{-31}$. More general choices of $w$ or multiple stages of inflation were studied in Refs.~\cite{ Mizuno:2019bxy,  Dhuria:2019oyf, Torabian:2019zms}.

Later, we will introduce relations between $N_\text{total}$ and $N_\text{CMB}$.  The latter is defined as the e-folding number between the horizon exit of the CMB modes and the end of inflation. The CMB scale is defined as the pivot scale $k_* = 0.05 \, \text{Mpc}^{-1}$~\cite{Akrami:2018odb}, so its relation to the horizon scale is given by
\begin{align}
N_\text{hor} - N_\text{CMB} = \log \frac{k_*}{H_0} \approx 5.4.
\end{align}

\section{TCC and the shape of inflaton potential} \label{sec:main}

Let us discuss the possible shape of the inflaton potential.
Because the TCC has to be satisfied for any initial conditions~\cite{Bedroya:2019snp}, an immediate consequence is that we cannot have a plateau-type potential.  In particular, the small $\alpha$ limit of $\alpha$-attractor models of inflation~\cite{Ellis:2013nxa, Ferrara:2013rsa, Kallosh:2013yoa, Kallosh:2014rga, Galante:2014ifa, Carrasco:2015pla, Roest:2015qya, Linde:2015uga, Scalisi:2015qga,Akrami:2017cir}, which can easily realize a tiny tensor-to-scalar ratio $r \propto \alpha$, is forbidden unless the shift symmetry of inflaton is broken.   This is because if the inflaton's initial value is on top of the plateau with a negligible velocity, it easily exceeds the e-folding number allowed by TCC.

The remaining possibility for the small-field inflation with a tiny $\epsilon$ is either the hilltop-inflation type~\cite{Linde:1981mu, Albrecht:1982wi, Boubekeur:2005zm} or the inflection-point-inflation type~\cite{Allahverdi:2006iq, Allahverdi:2006we, BuenoSanchez:2006rze, Baumann:2007np, Baumann:2007ah}. 
In the case of hilltop inflation, the condition for eternal inflation was studied in Ref.~\cite{Barenboim:2016mmw} by assuming a simple constant-plus-monomial potential, $V = V_0 (1 - (\phi/\mu )^p )$ with $p$ even.  It was shown that there always exists an initial condition for the eternal inflation to be induced around the top of the potential under the condition that $\mu$ is chosen to fit $n_\text{s} \simeq 0.96$.  Of course, eternal inflation is inconsistent with the TCC. 
It is remarkable that almost all (if not all) the universality classes of inflation~\cite{Mukhanov:2013tua, Roest:2013fha, Garcia-Bellido:2014gna, Binetruy:2014zya} are incompatible with the TCC.

Below, we consider inflaton potentials which have a limited flat region and an inflection point in two complementary approaches.
In Sec.~\ref{sec:expansion}, we take a ``model-independent'' phenomenological description in which we expand the inflaton potential as Taylor series around the CMB scale. In this approach, it is clear that the total e-fold can be as twice as one necessary for the CMB as long as the expansion approach itself is meaningful. Also, a qualitative argument shows a need for a sizable cubic term, which corresponds to a sizable negative running spectral index.  Some ${\cal O}(1)$ uncertainties are unavoidable in this approach, and we move to a more concrete approach in Secs.~\ref{sec:quartic} and \ref{sec:alpha-attractor}. 
 We will see that analyses in such concrete setups support our two main claims: a sizable negative running spectral index and an $\mathcal{O}(100 \%)$ tighter lower bound on the maximally possible e-folding number (as we vary the initial conditions).

\subsection{Expansion around the CMB scale} \label{sec:expansion}

Let us expand the inflaton potential around the CMB pivot scale, $\phi = \phi_\text{CMB} +  \varphi$,
\begin{align}
V = V_0 \left( 1 +  \sqrt{2\epsilon_0 } \varphi +  \frac{1}{2} \eta_0 \varphi^2 + \frac{1}{3} \mu \varphi^3 + \dots  \right),  \label{V_expansion}
\end{align}
where we shifted the coordinate origin so that $\varphi = 0$ corresponds to the CMB scale,  $V_0 < 6.4 \times 10^{-39}$ and $\epsilon_0 < 1.3 \times 10^{-32}$ satisfy the TCC constraint, and $\eta_0 \simeq - 0.02$ is the value of the second slow-roll parameter $\eta = V''/V$ which fits the spectral index $\ns = 1 - 6\epsilon + 2\eta \simeq 1 + 2 \eta$, whose observational value is $\ns = 0.9649\pm 0.0042$ (68\% confidence level (CL); Planck 2018 TT,TE,EE$+$lowE$+$lensing)~\cite{Akrami:2018odb}.  
Our convention is that inflation starts from a positive $\varphi$ and it rolls down in the negative direction.
Here, we do not aim to explain the values of the slow-roll parameters $\epsilon_0$ and $\eta_0$  at the pivot scale.  They appear in the coefficients just by construction in this phenomenological description. 
As mentioned above, the pure quadratic potential leads to eternal inflation, so the TCC necessitates the higher-order terms.

First, let us simply assume that the higher-order terms are negligible until the point where the linear and quadratic terms become comparable. (We will shortly see that we have to loosen this assumption.) When we compare two terms, we compare their contributions to the slope (first derivative) of the potential,
\begin{align}
V' = V_0 \left(  \sqrt{2\epsilon_0} + \eta_0 \varphi + \mu \varphi^2 + \dots \right), \label{V'_expansion}
\end{align}
since the slow-roll parameter $\epsilon (\varphi)$ is the relevant quantity.
  In this case, the field range where the linear approximation is valid in this description is 
\begin{align}
\Delta \varphi \simeq 2 \times \frac{ \sqrt{2\epsilon_0}}{|\eta_0|} \simeq 100 \sqrt{2 \epsilon_0} <  2 \times 10^{-14}, 
\end{align}
where $2$ has been multiplied to take into account both $\varphi>0$ and $\varphi<0$ regions. 
This factor $2$ is trivial in this example, but we emphasize that it is expected model-independently as long as the expansion around the CMB point makes sense. 
In this range, the slope is dominated by the linear term.
The possible e-folding number during slow-roll in this whole range is estimated as
\begin{align}
\frac{\Delta\varphi}{\partial \varphi / \partial N} \simeq \frac{2}{|\eta_0|} \simeq 100,
\end{align}
where we have used $\partial \varphi / \partial N = V'/V = \sqrt{2 \epsilon_0}$. 
This is typically larger than the TCC bound and marginally larger than that even if we take the minimally allowed inflation energy consistent with Big-Bang nucleosynthesis.

Of course, it is a nontrivial question of how to realize such a maximum possible slow-roll, which is essentially the initial condition problem of small-field inflation models. This is a separate issue distinguished from the TCC.  However unnatural the initial condition is, it needs to be consistent with the TCC condition.  
Therefore, finding one possible e-folding number in a model gives rise to a lower bound on the maximum possible total e-folding number $N_\text{total}^{\text{max}} = \displaystyle \max_{\text{initial conditions}} N_\text{total}$ (while the total number is upper bounded by the TCC).

To fix the above problem, we need to have a somewhat larger cubic term with a positive  $\mu$.    This makes the slopes in both  $\varphi>0$ and $\varphi<0$ regions steeper.  If the sign is opposite, there is still a hilltop in the $\varphi > 0$ region, and also the $\varphi < 0$ region is flatter, increasing the e-folding number.  On the other hand, if the cubic term is suppressed by some reasons and the quartic term dominates over the cubic term, it is not efficient to make it consistent with the TCC.  For example, $V_0 \lambda \varphi^4$ in the potential with $\lambda>0$ makes the slope flatter in the $\varphi < 0$ region, and $\lambda <0$ does not eliminate the hilltop in the $\varphi > 0$ region.
In principle, one could rely on higher odd-order terms like $\phi^5$ or $\phi^7$, but in which case one has to explain why all the lower order terms are suppressed.  Thus, in simple generic cases, a positive cubic term is necessary. 

Just before the $\varphi$ reaches the point where the linear and cubic term contributions are comparable, the order of magnitude of $\epsilon$ is still same as that at the CMB scale.  Therefore, we assume that all higher-order terms conspire to add up and end inflation quickly after they become comparable.  This corresponds to a conservative estimate on the amount of e-folding.  In this regime, our expansion approach completely breaks down. Still, working with the linear term before going into such a dangerous regime is a self-consistent working assumption~\cite{Bedroya:2019tba}. 

To satisfy $\frac{\Delta\varphi}{\partial \varphi / \partial N} = N_\text{total}$ for a given e-folding number $N_\text{total}$, we take
\begin{align}
\mu \simeq  \frac{4}{\sqrt{2\epsilon_0}N_\text{total}^2}.
\end{align}
Here, the notation $N_\text{total}$ implies that we assume that inflation ends soon after the cubic term (and other higher-order terms) become relevant and the linear approximation breaks down, as discussed in the previous paragraph.
In this case, the total e-folding number is nothing but twice as the e-folding experienced during inflation in $\varphi<0$ region,
\begin{align}
N_\text{total} \simeq 2 N_\text{CMB}.
\end{align}

Note that the cubic term is related not only to $N_\text{total}$ but also to the running spectral index.
The third slow-roll parameter $\xi^2 = V' V'''/ V^2$ is evaluated as $\xi^2 =  2 \sqrt{2 \epsilon_0}  \mu $.  The running of the spectral index $\alpha_\text{s}= 16 \epsilon \eta - 24 \epsilon^2 - 2\xi^2$ is negative,
\begin{align}
\alpha_\text{s} \simeq - 4 \sqrt{2 \epsilon_0} \mu \simeq  - \frac{16}{N_\text{total}^2} .
\end{align}
The Planck 2018 TT,TE,EE+lowE+lensing constraint is  $\alpha_\text{s} = -0.0045 \pm 0.0067$ at 68\% CL~\cite{Akrami:2018odb}. This gives an approximate constraint on $N_\text{total}$ as follows,
\begin{align}
N_\text{total} \gtrsim 38.  
\end{align} 
The future prospect of the precision of $\alpha_\text{s}$ measurement is $3 \times 10^{-4}$~\cite{Kohri:2013mxa}. If the central value is unchanged, the constraint becomes $N_\text{total} > 58$. 
In this case, the upper bounds on the inflation scale and the tensor-to-scalar ratio become $V_\text{inf}^{1/4} < \mathcal{O} (10^6 )\, \text{GeV}$ and $r < \mathcal{O}(10^{-42})$, respectively.

Note that the above discussion neglects ${\cal O}(1)$ coefficients for the estimation of $\mu$, whose uncertainty exponentially propagates to the upper bound on $V$ and $r$. 
Nevertheless, this analysis implies that, for a given inflation model, one has to be careful about the constraint from the running of the spectral index\footnote{In \cite{Kinney:2014jya}, it was pointed out 
that the present bound on the running spectral index is sufficient to prevent eternal inflation.  The relation between the short e-folding number and the running spectral index was also discussed in Ref.~\cite{Takahashi:2019qmh}.}.

In fact, we will see that stronger bounds on $\alpha_\text{s}$ are obtained, and longer total e-folding numbers ($N_\text{total} > 2 N_\text{CMB}$) are obtained, in the following subsections.  The main reason for such a difference is due to the fact that we assumed in this subsection that inflation ends when higher-order terms become comparable to the linear term, but the slow-roll conditions can be maintained for a while after this point in the inflation models studied below.  In this sense, the estimates in this subsection can be regarded as conservative.

\subsection{Concrete model analysis 1: Renormalizable quartic potential} \label{sec:quartic}
To discuss the implications of the TCC more concretely, let us consider an example model.
We begin with the renormalizable (quartic) potential
\begin{align}
V = & \frac{1}{2} m^2 \phi^2  - \frac{1}{3} \mu \phi^3 + \frac{1}{4} \lambda \phi^4.
\end{align}
We required that the cosmological constant vanishes at the vacuum $\phi=0$, i.e.~$V(\phi=0) = 0$. 
We exchange the parameters ($m^2$, $\mu$, $\lambda$)  to a different set ($\phi_0$, $\Delta$, $\lambda$):  $m^2 \equiv  \lambda ( \phi_0^2 + \Delta^2)$ and  $\mu \equiv 2 \lambda \phi_0$, which will be useful below.  We take $\phi_0 > 0$ as a convention. This parametrization is designed so that the first derivative of the potential has the form
\begin{align}
V' = & \lambda \phi \left( (\phi - \phi_0)^2 + \Delta^2  \right).
\end{align}
That is, the slope vanishes at $\phi=0$ and $\phi = \phi_0$ in the limit $\Delta =0$.
The point $\phi = \phi_0$ is also an inflection point in the limit $\Delta =0$.
The model in this limit is known as the renormalizable inflection point inflation model~\cite{Martin:2013tda, Allahverdi:2006cx, Allahverdi:2007wt}.
The small parameter $\Delta$ is introduced to tune the shape of the potential.
Note that  eternal (or at least very long) inflation can occur at $\phi = \phi_0$ in the absence of $\Delta$, so we definitely need a nonzero $\Delta$.

For our analyses, it is convenient to redefine the origin of the field, $\phi = \varphi + \phi_0$, and make the field $\varphi$ and the mass dimensional parameter $\Delta$ dimensionless: 
\begin{align}
\varphi = \widetilde{\phi} \phi_0, \quad 
\Delta = \widetilde{\Delta} \phi_0.
\end{align}
Variables with a tilde are dimensionless.
The potential now reads
\begin{align}
V = \frac{\lambda \phi_0^4}{12}  \left( 1 + \widetilde{\phi} \right)^2 \left( 1- 2 \widetilde{\phi} + 3 \widetilde{\phi}^2 + 6 \widetilde{\Delta}^2 \right).
\end{align}

By construction, inflation occurs at a region close to $\widetilde{\phi} = 0$ ($\phi = \phi_0$), and $\widetilde{\Delta}$ is also introduced as a tiny parameter.  Therefore, we self-consistently assume $|\widetilde{\phi}| \ll 1$ during inflation and $\widetilde{\Delta} \ll 1$.  

First, let us discuss the e-folding number.  Using the slow-roll formula,
\begin{align}
\frac{1}{\sqrt{2 \epsilon}} = \frac{\phi_0 (1 + \widetilde{\phi}) ( 1-2 \widetilde{\phi} + 3 \widetilde{\phi}^2 + 6 \widetilde{\Delta}^2 )}{12 (\widetilde{\phi}^2 + \widetilde{\Delta}^2)} .
\end{align}
Note that the dominant contribution to the e-folding number comes from the region where $\widetilde{\phi}$ is at most comparable to $\widetilde{\Delta}$ so that the denominator gets suppressed.  As anticipated above, for such a region, it is a good approximation to take $|\widetilde{\phi} | \ll 1$, which gives $1/\sqrt{2\epsilon} \simeq \phi_0 / 12(\widetilde{\phi}^2 + \widetilde{\Delta}^2 )$.
Using this, the e-folding number is obtained as
\begin{align}
 N = \left [  \frac{\phi_0^2 \arctan (\widetilde{\phi} /\widetilde{ \Delta} )}{12 \widetilde{\Delta}}  \right]^{\widetilde{\phi}=\widetilde{\phi}_\text{begin}}_{\widetilde{\phi} = \widetilde{\phi}_\text{end}}.
\end{align}
Note that $\arctan (x)$ grows linearly as $x$ from $\arctan(0) = 0$, quickly reaches a half of maximum value at $x = 1$ as $\arctan(1) = \pi / 4$, and then saturates soon, e.g.~$\frac{\arctan(10)}{\lim_{x\to\infty}\arctan(x)} = 93\%$, and finally asymptote to the final value  $\lim_{x \to \infty} \arctan (x) = \pi/ 2$.
This means that the substantial e-folding number comes only from the region around $\widetilde{\phi} \lesssim \mathcal{O}( \widetilde{\Delta})$.
The asymptotic value of the total available e-folding number in the slow-roll region is
\begin{align}
N_\text{total} = \frac{\pi \phi_0^2}{12 \widetilde{\Delta}},
\end{align}
where we have considered both $\widetilde{\phi}<0$ and $\widetilde{\phi}>0$ regions.
Inverting this, one obtains $\widetilde{\Delta} = \pi \phi_0^2 / 12 N_\text{total}$.

Next, we consider the second slow-roll parameter $\eta$ to explain the spectral index $\ns$.
At the CMB scale, $\eta$ can be approximated as
\begin{align}
\eta =& \frac{12 (2 \widetilde{\phi}  + 3 \widetilde{\phi} ^2 + \widetilde{\Delta}^2)}{\phi_0^2 (1 + \widetilde{\phi})^2 (1 - 2 \widetilde{\phi} + 3 \widetilde{\phi} ^2 + 6 \widetilde{\Delta}^2 )} \simeq \frac{24 \widetilde{\phi} }{\phi_0^2},
\end{align}
where we assumed $1 \gg |\widetilde{\phi}| \gg \widetilde{\Delta}^2$ at the CMB scale.  Numerically, this assumption is well satisfied in the parameter region of our interests. 
For a given input value of $\eta$, one obtains the field value at the CMB scale as $\widetilde{\phi} = \eta \phi_0^2 / 24$.

The first slow-roll parameter at the CMB scale can be fit by $\phi_0$ by solving
\begin{align}
\epsilon \simeq \frac{\phi_0^6 (\eta^2 N_\text{total}^2 + 4\pi^2)^2}{4608 N_\text{total}^4},
\end{align}
where we have used $\widetilde{\Delta} = \pi \phi_0^2 / 12 N_\text{total}$ and $\widetilde{\phi} = \eta \phi_0^2 / 24$.

Finally, the relation between $N_\text{CMB}$ and $N_\text{total}$ can be found by solving
\begin{align}
N_\text{CMB} = & \frac{N_\text{total}}{2} - \frac{\phi_0^2 }{12 \widetilde{\Delta}} \arctan \left( \frac{|\widetilde{\phi}|}{\widetilde{\Delta}} \right) \nonumber \\
=& \frac{N_\text{total}}{2} \left( 1 - \frac{2}{\pi} \arctan \left( \frac{|\eta| N_\text{total}}{2\pi} \right)  \right).  \label{N_CMB_quartic}
\end{align}

We also calculate the running spectral index $\alpha_\text{s}$.
Using the same approximation as above valid at the CMB scale, we have the third slow-roll parameter
\begin{align}
\xi^2 = & \frac{288 (1 + 3 \widetilde{\phi}) (\widetilde{\phi}^2 + \widetilde{ \Delta}^2)}{\phi_0^4 (1 + \widetilde{\phi})^3 (1 - 2 \widetilde{\phi} + 3 \widetilde{\phi} ^2 + 6 \widetilde{\Delta}^2 )}  
\simeq   \frac{\eta^2}{2} + \frac{2\pi^2}{N_\text{total}^2} ,
\end{align} 
where the term proportional to $\eta^2$ is numerically subdominant, and this gives
\begin{align}
\alpha_\text{s} \simeq  - \frac{4 \pi^2}{N_\text{total}^2}.  \label{alpha_s_quartic}
\end{align}

For the consistency check of $|\widetilde{\phi}| \gg \widetilde{\Delta}^2$, we take the ratio
\begin{align}
\frac{|\widetilde{\phi}|}{\widetilde{\Delta}^2} = \frac{6 |\eta| N_\text{total}^2}{\pi^2 \phi_0^2},
\end{align}
where $\eta$ is evaluated at the CMB scale.  This is large because the model is a small-field model $\phi_0 \ll 1$, so the above assumption is self-consistent.

\begin{figure}[htb]
 \centering
\includegraphics[width=0.6 \columnwidth]{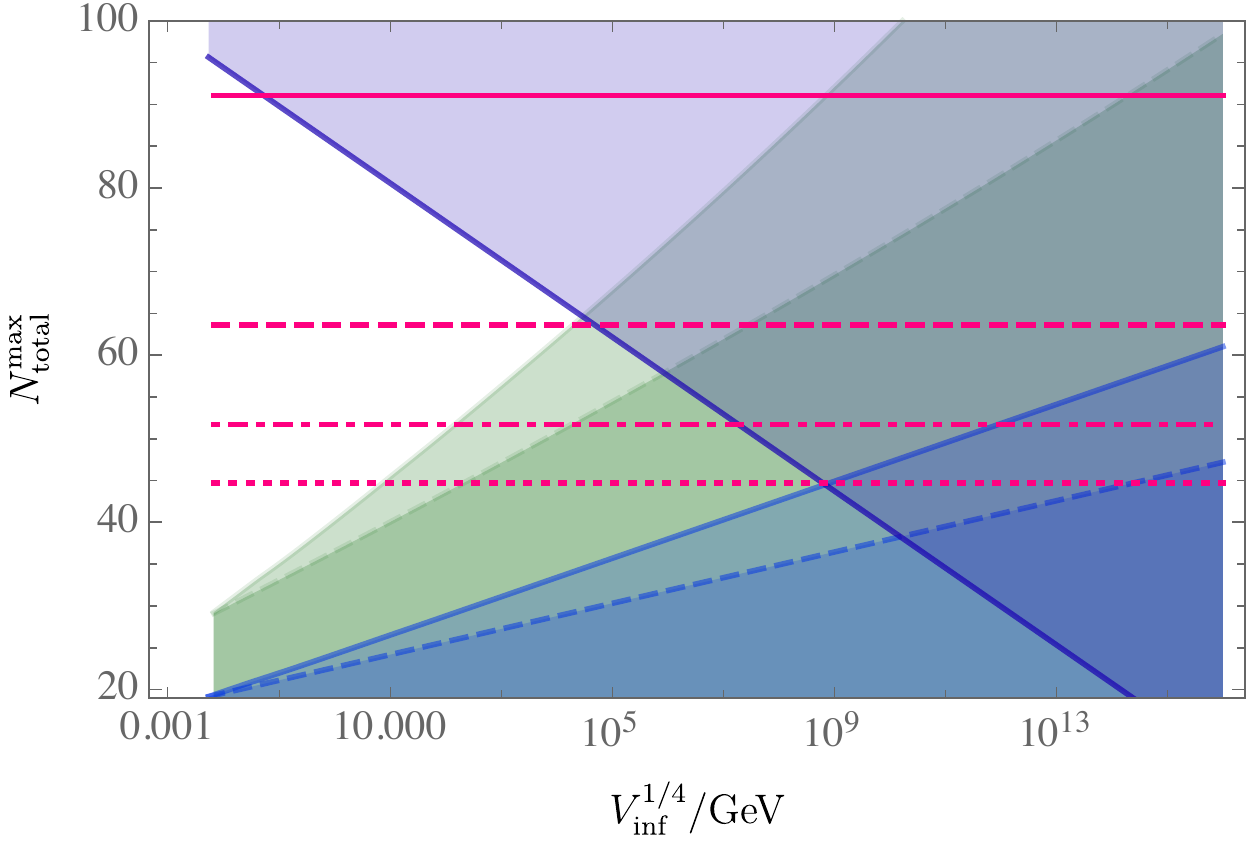}
  \caption{Constraints on the maximum total e-folding number $N_\text{total}^\text{max}$ as a function of the the inflationary energy scale $V_\text{inf}^{1/4}$. The  purple shaded region is excluded from the TCC constraint, \eqref{TCC}.  The blue shaded region is excluded by the horizon problem, \eqref{horizon}, (Solid: the highest possible $T_\text{R} = (30 V_\text{inf}/\pi^2 g_{*}(T_\text{R}))^{1/4} $; Dashed: the lowest possible $T_\text{R} = 4 \, \text{MeV}$).  The green region is excluded by the additional e-folding number (e-folds before the CMB scale mode exits the horizon) discussed in this paper under the condition that $\ns$ is consistent with the Planck data [see eq.~\eqref{N_CMB_quartic}] (Solid: the highest possible $T_\text{R} = (30 V_\text{inf}/\pi^2 g_{*}(T_\text{R}))^{1/4} $; Dashed: the lowest possible $T_\text{R} = 4 \, \text{MeV}$). The intersection of the purple line and the green line gives eq.~\eqref{Vinf_upper-bound_quartic}.  The solid, dashed, dot-dashed, and dotted magenta lines are lower bounds on $N_\text{total}^\text{max}$ when the lower bound on  $\alpha_\text{s}$ is given as   -0.005, -0.010, -0.015,  and -0.020, respectively [see eq.~\eqref{alpha_s_quartic}].
}
 \label{fig:N_range_quartic}
  \end{figure}

For a given inflationary energy scale $V_\text{inf}$, $N_\text{CMB}$ is determined from the requirement to solve the horizon problem under an assumption of reheating history.
Using $N_\text{CMB}$ and the $\ns$ data, we can deduce the total available e-folding number in the slow-roll region $N_\text{total}$ from eq.~\eqref{N_CMB_quartic}. This gives a lower bound on the maximum total e-folding number $N_\text{total}^\text{max} = \displaystyle \max_{\text{initial conditions}} N_\text{total}$.  Such a constraint is shown in Fig.~\ref{fig:N_range_quartic}.  For degrees of freedom, we assume that of the Standard Model and used the tabulated data given by Ref.~\cite{Saikawa:2018rcs}.

  In the figure, we also draw the potential lower bound on $N_{\text{total}}^{\text{max}}$ assuming that the lower bounds on $\alpha_\text{s}$ are obtained as -0.005, -0.010, -0.015, and -0.020, corresponding to the solid, dashed, dot-dashed, and dotted magenta lines, respectively.
    The lines are horizontal since $\alpha_\text{s}$ and $N_\text{total}$ are in one-to-one correspondence as in eq.~\eqref{alpha_s_quartic}. The relation is insensitive to the inflation energy scale or $N_\text{CMB}$ because the running is essentially contributed by the cubic term around the inflection point, i.e.~the dependence on $\widetilde{\phi} = \widetilde{\phi}(N_\text{CMB})$ shows up in subdominant terms. 
  If we take $\alpha_\text{s} \gtrsim - 0.004$, there is no allowed region.  Thus, the TCC puts an upper  bound on the value of the running spectral index in this model,
  \begin{align}
  \alpha_\text{s} \lesssim -4.0 \times 10^{-3}.
  \end{align}

Because of the possible additional e-folding number (the additional e-folds here refers to the number of e-folds before the CMB scale mode exists the horizon), the upper bounds on the inflation energy and the tensor-to-scalar ratio become tighter 
\begin{align}
V_\text{inf}^{1/4} <& 
3.0 \times 10^4 \, \text{GeV}  , 
&
r < & 
7.2 \times 10^{-49}  ,  \label{Vinf_upper-bound_quartic}
\end{align}
where the instantaneous reheating is assumed.  If we consider a matter-dominated era ending with the reheating temperature $T_\text{R} = 4\, \text{MeV}$~\cite{Hannestad:2004px, deSalas:2015glj, Hasegawa:2019jsa}, the constraints become $V_\text{inf}^{1/4} < 8.7 \times 10^5 \, \text{GeV}  $ and $r < 5.1 \times 10^{-43}$.

\subsection{Concrete model analysis 2: $\alpha$-attractor with broken shift symmetry} \label{sec:alpha-attractor}

To reproduce the tiny $\epsilon$, the $\alpha$-attractor~\cite{Ellis:2013nxa, Ferrara:2013rsa, Kallosh:2013yoa, Kallosh:2014rga, Galante:2014ifa, Carrasco:2015pla, Roest:2015qya, Linde:2015uga, Scalisi:2015qga,Akrami:2017cir} may be a good starting point.
Of course, it is a plateau type model, so the shift symmetry of the inflaton must be broken, such that the plateau has only a finite field range not to conflict with the TCC. For this type of model, the asymptotic ($\phi \gg \sqrt{3\alpha /2}$) behavior of the potential is generically
\begin{align}
V = V_0  \sum_{n \geq 0}\left( c_n e^{-n \sqrt{\frac{2}{3\alpha}}\phi} + d_n \varepsilon _{\text{br}}^{n} e^{ n \sqrt{\frac{2}{3\alpha}}\phi}  \right),  \label{V_exp-expansion}
\end{align}
where $\alpha >0$ is a parameter, $c_m$ and $d_n$ are dimensionless coefficients of order unity, and $\varepsilon_{\text{br}} \ll 1$ is a symmetry breaking parameter, whose smallness is technically natural~\cite{tHooft:1979rat}.
Without the breaking terms ($\varepsilon_{\text{br}}=0$), the prediction of the $\alpha$-attractor is given by 
\begin{align}
  r=& \frac{12 \alpha }{N^2}, &  \ns -1=& - \frac{2}{N}, &   \alpha_\text{s} =& - \frac{2}{N^2}.
\end{align}
One can see that $r$ becomes arbitrarily small in the limit $\alpha \to 0$.  Also, in the limit $\varepsilon_{\text{br}} =0$, eq.~\eqref{V_exp-expansion} can be viewed as an expansion around $\phi\to \infty$ where the shift symmetry becomes better and better. Quantum consistency of such a potential is shown in Ref.~\cite{Bezrukov:2010jz}. This expansion can also be interpreted as a power series expansion of the potential in the defining frame (before field redefinition) 
  of $\alpha$-attractor models.  On top of such a potential, we added technically natural breaking terms, which is a power series with respect to $\varepsilon_{\text{br}}$.

Let us consider the potential
\begin{align}
V = V_0 \left( 1 - (1+\varepsilon_{\text{br}} ) e^{-\sqrt{\frac{2}{3\alpha}}\phi} + \varepsilon_{\text{br}} e^{\sqrt{\frac{2}{3\alpha}}\phi} \right)^2 ,  \label{V_alpha-model}
\end{align}
where $\varepsilon_{\text{br}} (>0)$ in the second term is just to ensure $V(\phi=0)= 0$ is the minimum. 
The last term is responsible for the exponential rising of the potential which breaks the plateau.
This kind of rising of the potential (and the power suppression of the curvature perturbations at low multipoles) has been discussed in the literature in various contexts; see e.g.~Refs.~\cite{Broy:2014xwa, Abe:2014opa, Broy:2015qna, Asaka:2015vza}.
The above form of the potential sets the global property of the potential, e.g.~there are no local minima. 
For our purposes to analyze inflationary quantities, we can safely neglect the higher-order terms.
Thus, we work with
\begin{align}
V =& V_0 \left( 1 - 2 e^{-\lambda \phi} +  2 \varepsilon_{\text{br}} e^{ \lambda \phi} \right),
\end{align}
which keeps the leading terms in the expansion in eq.~\eqref{V_exp-expansion} and matches eq.~\eqref{V_alpha-model}.   We also introduced a new parameter $\lambda \equiv \sqrt{2/3\alpha}\, (\gg 1)$ for a shorthand notation.

The slow-roll parameters are 
\begin{align}
\epsilon = & 2 \lambda^2 \left( e^{-\lambda \phi} +  \varepsilon_{\text{br}} e^{\lambda \phi} \right)^2, \\ 
\eta = & 2 \lambda^2 \left( - e^{-\lambda \phi} +  \varepsilon_{\text{br}} e^{\lambda \phi} \right), \\ 
\xi^2 = & 4 \lambda^4 \left( e^{-\lambda \phi} +  \varepsilon_{\text{br}} e^{\lambda \phi} \right)^2.
\end{align}
where we neglected terms higher oder in $e^{- \lambda \phi}$ and in $\varepsilon_{\text{br}} e^{\lambda \phi} $.
The e-folding number in the slow-roll regime is
\begin{align}
N = \left [ \frac{\arctan [ \sqrt{\varepsilon_{\text{br}}}   e^{\lambda \phi} ] }{2 \sqrt{ \varepsilon_{\text{br}}} \lambda^2 } \right ]^{\phi = \phi}_{\phi = \phi_\text{end}},  \label{N_formula_1}
\end{align}
where $\phi_\text{end}$ is determined by $\eta = -1$, where slow-roll formulas become invalid.  In our setup, $\epsilon$ is still much smaller than unity when this happens, but the e-folds between the end of slow-roll and the end of inflation (the inflation could persist without satisfying the slow-roll conditions) is small because of the rapidly increasing inflaton velocity.  
The above e-folding expression reduces to $e^{\lambda \phi}/(2\lambda^2)$ in the limit of $\varepsilon_{\text{br}} \to 0$. 

To discuss the maximum total e-folding number, imagine that the upper-end value of $\phi$ in eq.~\eqref{N_formula_1} is increased.
When $\phi$ becomes sufficiently large that the correction term becomes relevant, it becomes hard to earn additional e-folding numbers.
As $\phi$ goes further, the validity of slow-roll ends at some point $\phi = \phi_\text{begin}$ where $\eta =1$.  We can extend the slow-roll analysis up to this point, so we consider the total e-folding number $N_\text{total}$ from this point to the point where (slow-roll) inflation ends.  
One may consider different initial conditions such that $\phi$ moves from much further point in the field space where slow-roll is not possible but inflation is still possible (for example $\epsilon \ll 1$ and $\eta \gg 1$), but the inflaton will gain a large velocity in this region, and it will quickly pass the flat region which can support slow-roll dynamics. (We consider another initial condition in Appendix~\ref{sec:rolling-up} in which inflaton first rolls up the potential and then rolls down, but the e-folding number during the rolling-up phase turns out to be small. The analyses in this Appendix also implies that the e-folds available from such a fast motion is limited.)  Therefore, we do not consider such initial conditions. We consider an initial condition in which $\phi$ begins to slow-roll the potential from $\phi = \phi_\text{begin}$, passes through the CMB point $\phi_\text{CMB}$, and reaches $\phi_\text{end}$.  As already mentioned in Sec.~\ref{sec:expansion}, how to realize such an initial condition (the initial condition problem of small-field inflation in general) is a separate issue.  We utilize the fact that  $N_\text{total}$ in any initial condition, however unnatural it is, gives a lower bound on $N_\text{total}^{\text{max}} = \displaystyle \max_{\text{initial conditions}} N_\text{total}$.  

It is useful to define a combination
\begin{align}
c \equiv & 4 \varepsilon_{\text{br}} \lambda^4 N_\text{CMB}^2,
\end{align}
which characterizes the size of the correction at the CMB scale.  Using this,  the total e-folding number can be expressed as
\begin{align}
N_\text{total} = \frac{N_\text{CMB}}{\sqrt{c}} \left( \arctan \left [  \frac{N_\text{CMB}}{\sqrt{c}}   \right] - \arctan \left [ \frac{\sqrt{c}}{N_\text{CMB}} \right ] \right).  \label{N_total}
\end{align}
The second term is approximately unity because of $\arctan x = x + \mathcal{O}(x^3 )$. Thus, the total e-folding number can be written in terms of $N_\text{CMB}$ and $c$.  Now, a key fact is that if we assume too large a value of $c$ to shorten the total e-folding number, it strongly affects the CMB observables such as $\ns$ and $\alpha_\text{s}$ because it modifies the shape of the potential significantly at the CMB scale.

To obtain the dependence of inflationary observables on $c$, we need to invert the relation \eqref{N_formula_1} between $N$ and $\phi$.  In our analysis, we do this recursively by the perturbation method with respect to $c$, and we retain up to the second order $c^2$.  The results are
\begin{align}
\lambda^2 e^{-\lambda \phi_\text{CMB}} =& \frac{1}{2N_\text{CMB}} \left( 1 - \frac{c}{3} - \frac{c^2}{45} + \mathcal{O}(c^3)\right)  , \\ 
\lambda^2 \varepsilon_{\text{br}} e^{\lambda \phi_\text{CMB}} =& \frac{1}{2 N_\text{CMB}} \left( c + \frac{ c^2}{3}  + \mathcal{O}(c^3)   \right) .
\end{align}
Using these pieces, we can calculate inflationary observables.
\begin{align}
r = &  \frac{12 \alpha }{N_\text{CMB}^2} \left( 1 + \frac{4c}{3} + \frac{16 c^2}{15} + \mathcal{O}(c^3)  \right) , \\
\ns -1 =& - \frac{2}{N_\text{CMB}}  \left( 1 -\frac{4c}{3} - \frac{16 c^2}{45} + \mathcal{O}(c^3)  \right) , \label{ns_alpha-attractor} \\
\alpha_\text{s} =& - \frac{2}{N_\text{CMB}^2} \left( 1 + \frac{4c}{3} + \frac{16 c^2}{15} + \mathcal{O}(c^3)  \right). \label{alphas_alpha-attractor}
\end{align}

\begin{figure}[htb]
 \centering
\includegraphics[width=0.6 \columnwidth]{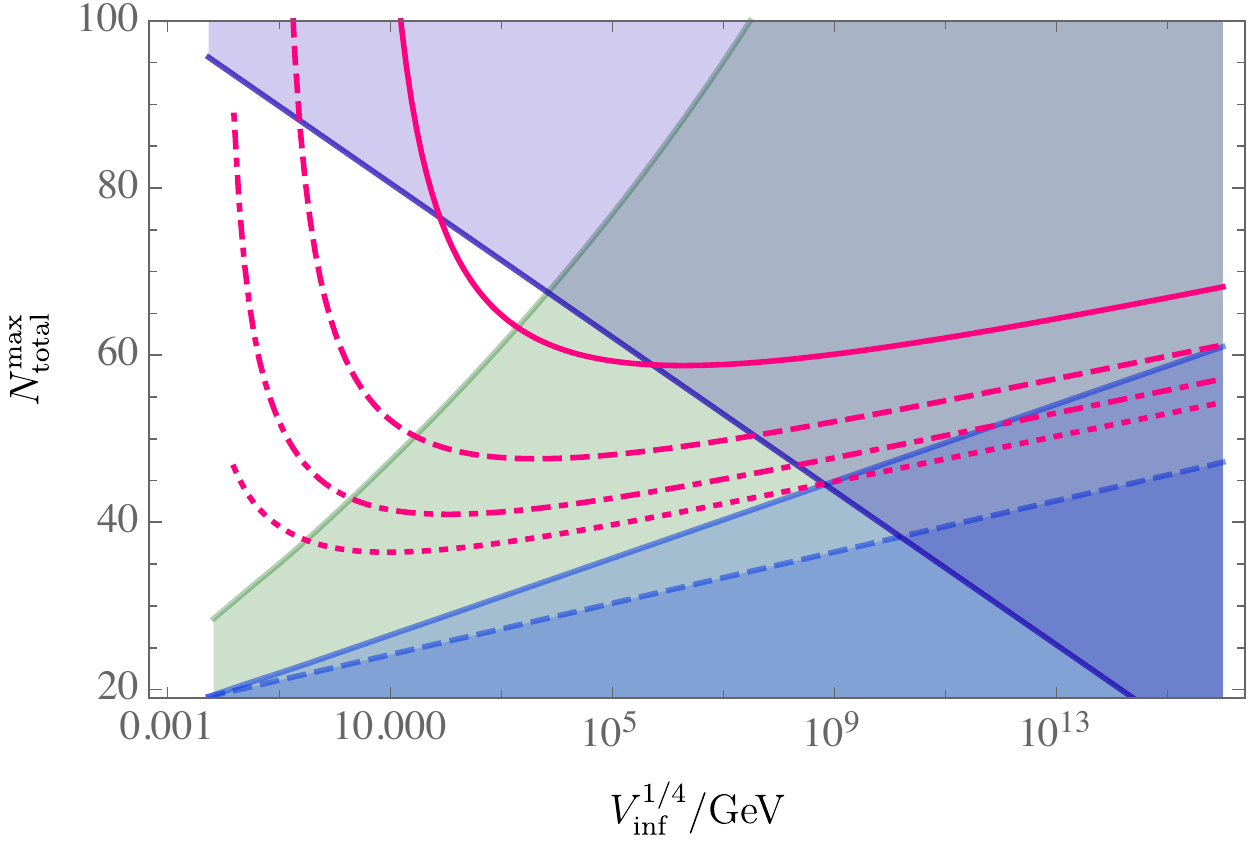}
  \caption{Constraints on the maximum total e-folding number $N_\text{total}^\text{max}$ as a function of the the inflationary energy scale $V_\text{inf}^{1/4}$. The  purple shaded region is excluded from the TCC constraint, \eqref{TCC}.  The blue shaded region is excluded by the horizon problem, \eqref{horizon}, (Solid: the highest possible $T_\text{R} = (30 V_\text{inf}/\pi^2 g_{*}(T_\text{R}))^{1/4} $; Dashed: the lowest possible $T_\text{R} = 4 \, \text{MeV}$).  The green region is excluded by the additional e-folding number discussed in this paper (under the condition that $\ns$ is consistent with the Planck data) [see eqs.~\eqref{N_total} and \eqref{ns_alpha-attractor}]. The intersection of the purple line and the green line gives eq.~\eqref{Vinf_upper-bound_alpha-attractor}. The solid, dashed, dot-dashed, and dotted magenta lines are the lower bounds on $N_\text{total}^\text{max}$ when the lower bound on  $\alpha_\text{s}$ is given as - 0.005, -0.0075, -0.010, and -0.0125, respectively [see eqs.~\eqref{N_total} and \eqref{alphas_alpha-attractor}]. The instantaneous reheating is assumed for the green line and magenta lines.
}
 \label{fig:N_range_n=1}
  \end{figure}

For a given inflationary energy scale $V_\text{inf}$, $N_\text{CMB}$ is fixed under the assumed reheating history.  We assume the instantaneous reheating as discussed in Sec.~\ref{sec:discussion}.  From the upper bound on $\ns = 0.9649\pm 0.0042$ 
  \cite{Akrami:2018odb}, we obtain an upper bound on the correction parameter $c$.  This gives a lower bound on $N_\text{total}^{\text{max}}$.  This strengthens the upper bounds on the inflationary energy scale and on the tensor-to-scalar ratio.

\begin{align}
V_\text{inf}^{1/4} <& 
7.0 \times 10^3 \, \text{GeV}   , 
&
r < &
2.2 \times 10^{-51}  . \label{Vinf_upper-bound_alpha-attractor}
\end{align}
  The allowed range of the total e-folding number is shown in Fig.~\ref{fig:N_range_n=1}. 
We also see that the running spectral index is upper bounded as 
 \begin{align}
 \alpha_\text{s} \lesssim - 4 \times 10^{-3},
\end{align}
to be consistent with the TCC.  If we obtain a lower bound like $ \alpha_\text{s} \gtrsim - 4 \times 10^{-3}$ from observations, the almost all parameter space for $N_\text{total}^\text{max}$ is excluded. 

It is interesting to note that the $c$ dependence of $\ns$ and $\alpha_\text{s}$ are opposite.
For the low energy scale inflation with a small $N_\text{CMB}$ (the left side of the figure), we need a sizable amount of the correction $c$ to fit $\ns$.  This means that the rising effect of the potential shape is sizable, so the bound on $N_\text{total}$ from $\ns$ (green shade) is weak. On the other hand, the absolute value of $\alpha_\text{s}$ becomes too large unless we take a small $c$. The small $c$ corresponds to a wide range of the plateau, which lets $N_\text{total}$ become large. This is the reason why the magenta lines sharply rise at the left side of Fig.~\ref{fig:N_range_n=1}.  For higher energy scale with a larger $N_\text{CMB}$ (the right side of the figure), $\ns$ can be explained without the correction $c$, so the allowed range of $c$ is small corresponding to a large $N_\text{total}$ (The height of the green shade becoming larger).  Meanwhile, the constraint on $c$ from $\alpha_\text{s}$ becomes weaker, so the gain $N_\text{total} - N_\text{CMB}$ becomes smaller for magenta lines.

  It is also worth noting in which part of the figure the perturbative treatment with respect to $c$ is better.
  As discussed in the previous paragraph, $c$ is smaller in the high energy side on the green line, but it is so in the low energy side on the magenta lines. 
Therefore, the convergence of the perturbative series in $c$ is better in the high energy inflation side for $\ns$ (green line), while it is better in the low energy inflation side for $\alpha_\text{s}$ (magenta lines).  The vertical distance between these lines and the blue solid line (more precisely, the ratio $N_\text{total}/N_\text{CMB}\sim 1/\sqrt{c}$) can be regarded as a measure of convergence of the perturbation series (see eq.~\eqref{N_total}).

One way to relax the above severe constraint is to assume that the correction term has a higher exponent, $\varepsilon_{\text{br}} e^{+n \lambda \phi}$ with $n>1$.  This is studied in Appendix~\ref{sec:high-order-breaking}.

The analysis made in this subsection is based on a particular model or expansion,  but we expect similar results for any inflation model which satisfies the following conditions:
\begin{itemize}
\item The inflaton potential (including the technically-natural shift-symmetry-breaking terms) is controlled by a systematic expansion.  
\item The first nonvanishing symmetry breaking term is not a very high order term (corresponding to $n = \mathcal{O}(1)$).
\item The dominant contribution to the e-folding number comes from the most flat part of the potential which is at the CMB scale or behind it $\phi_{\text{flattest}} \gtrsim \phi_\text{CMB} (> \phi_\text{end})$. 
\end{itemize}
When the third condition is not satisfied, our argument for $N_\text{total} \gtrsim 2 N_\text{CMB}$ is not applicable.  Even in this case, however, one should make sure that an initial condition with $\phi \simeq \phi_\text{flattest}$ and a negligible velocity does not lead to too long an e-folding number.

\section{Summary and Discussion} \label{sec:discussion}

We have discussed the implications of the TCC on the shape of the inflation potential focusing on single-field slow-roll models.  In the first ``model-independent'' approach, we have expanded the inflaton potential around the CMB pivot scale and argued that a sizable cubic term would be required to be consistent with the TCC.  This leads to a potentially strong lower bound on the e-folding number $N_\text{CMB}$.  Unfortunately, there is at least $\mathcal{O}(1)$ uncertainty for the estimation of the cubic term  since we do not know where exactly higher-order terms conspire to sum up to end inflation.  

For a more concrete discussion, we have taken a second approach in which we have considered the concrete models: the renormalizable quartic potential and the $\alpha$-attractor models of inflation with a shift symmetry broken in a controlled fashion.  Our analysis shows that, for such smooth and flat potentials, the e-folding number dominantly increases around (both before and after) the CMB pivot scale, which implies the total e-folding number must be at least doubled, $N_\text{total} \gtrsim 2 N_\text{CMB}$.  This strengthens the  constraints of the TCC on the energy scale of inflation.

Although there is an allowed region in Figs.~\ref{fig:N_range_quartic} and ~\ref{fig:N_range_n=1}, there remain  theoretical questions.
How can one protect the accidental flatness of the polynomial potential in Sec.~\ref{sec:quartic} from the radiative corrections?  For the radiative stability, the $\alpha$-attractor studied in Sec.~\ref{sec:alpha-attractor} is more advantageous.  However, we do not know what the physical interpretation or particle physics realization of the tiny $\alpha$ would be.  Is the ultraviolet completion of the model possible?  For example, the values derived in the maximal supergravity and M/String Theory in Ref.~\cite{Ferrara:2016fwe} are $3 \alpha =1, 2, \dots , 7$.  It is then unclear  whether or not the tiny value of $\alpha$, which might look ad hoc, belongs to the Landscape rather than the Swampland. 

A small $\alpha$ also introduces a strong coupling around the vacuum $|\phi| \lesssim \sqrt{\alpha}$ although it is weakly coupled during inflation. Thus, it is difficult to make a prediction on the (p)reheating dynamics,  but it is tempting to think the reheating completes quickly by such strong interactions with itself and with other fields.   Our assumption of instantaneous reheating is partially motivated by this strong coupling phenomenon.

Let us also briefly discuss the hybrid inflation~\cite{Linde:1993cn} where the single-field inflation trajectory ends with the instability of water-fall fields.
This mechanism is efficient to quickly end inflation, so some of the constraints discussed in this paper such as that on the running spectral index can be circumvented.  We could also impose a water-fall instability at the top of the hilltop inflaton trajectory for the eternal or very long inflation not to occur.  Still, it is a nontrivial challenge to suppress quantum corrections from loops of water-fall fields to obtain a tiny slow-roll parameter.

Although the main concern in this paper is the TCC, we briefly discuss other Swampland conjectures too.  There has been accumulating evidence on the difficulties to realize dS spacetime in the asymptotic region of the moduli space.  The most nontrivial point of the dS conjecture appears when one extends such observations to all the regions of the field space.  Suppose now that the constraints~\eqref{dS} indeed applies to the small-field inflaton potential.  
The first inequality of~\eqref{dS} is in sharp contradiction with one of the consequences of the TCC, $\epsilon \lesssim \mathcal{O} (10^{-51})$ (see inequalities~\eqref{Vinf_upper-bound_quartic} and \eqref{Vinf_upper-bound_alpha-attractor}).  At the CMB scale, the value of $\eta$ leads to $|c'| < 1.965 \times 10^{-2}$.  More importantly, however, our discussion implies that there should be an (approximate) inflection point where $V''$ vanishes.  Because of the smallness of $\eta$, we do not expect a rapid increase of $\epsilon$, so both inequalities are violated around the inflection point unless there is another unstable direction in the field space as discussed in the previous paragraph.  If this is taken seriously, one should consider models beyond the single-field slow-roll inflation.

On the other hand, the TCC is consistent with the distance conjecture~\eqref{distance}.  The estimate in Ref.~\cite{Bedroya:2019tba} already showed that the typical field distance during inflation is about $|\Delta \phi| < 10^{-13}$.  In our examples, the maximum of $\Delta \phi$ during inflation is of order $10^{-22}$ and $10^{-26}$ in Secs.~\ref{sec:quartic} and \ref{sec:alpha-attractor}, respectively.

One good aspect of our findings is that the TCC can be tested by the future precise measurements of $\alpha_\text{s}$ with precision $3 \times 10^{-4}$~\cite{Kohri:2013mxa}.  Also, our new upper bounds on the inflationary energy scale, which is around 10~TeV, might imply possibilities that the physics of inflation is related to the hierarchy problem of the Higgs mass and accessible by the colliders.

\section*{Acknowledgment} \label{sec:ack}
TT thanks Ayuki Kamada and Kazunori Kohri for discussions.
This work was supported by IBS under the project code, IBS-R018-D1.

\appendix

\section{Additional e-folding from rolling up the potential} \label{sec:rolling-up}

In this appendix, we consider a reversed trajectory of single-field slow-roll inflation, and estimate the number of the additional e-folding $N_\text{up}$.  Even though such an initial condition may not be necessarily natural, we have to consider such an initial condition to study the TCC since the TCC is a requirement for all physically possible initial conditions~\cite{Bedroya:2019snp}.

During the roll-up phase, the slope term in the equation of motion is negligible compared to the acceleration and the Hubble friction.
\begin{align}
\ddot{\phi} + 3 H \dot{\phi} \simeq 0.
\end{align}
This is the so-called ultra-slow-roll (USR) regime of inflation~\cite{Tsamis:2003px, Kinney:2005vj, Namjoo:2012aa, Martin:2012pe}.  The kinetic energy of inflaton is sufficiently large so that the slope of the potential is negligible, but it can still be much smaller than the potential energy.  This condition is $\frac{1}{2}\dot{\phi}(0)^2 < V$, where $t=0$ is the beginning of the USR phase. 
The above equation of motion is integrated to give
\begin{align}
\phi (t) = - \frac{\dot{\phi}(0)}{3 H} e^{-3Ht} + \phi (\infty ).
\end{align}
We regard approximately $H$ and $V$ as constants, and $Ht = N_\text{up}$ follows.

For simplicity, we approximate it as the USR regime from $t=0$ to $t=t_\text{turn}$ and the slow-roll regime after $t = t_\text{turn}$. The USR regime ends because the acceleration term and the slope term become comparable, $|\ddot{\phi}(t_\text{turn})| \sim V'  = V \sqrt{2 \epsilon_\text{turn}}$. 
This gives
\begin{align}
|\dot{\phi}(0) | = \frac{\sqrt{2 \epsilon_\text{turn}} V}{3 H} e^{3 N_\text{up}}.
\end{align}

We impose two conditions on the above quantity.
The first one is the condition that the initial kinetic energy must be smaller than the potential, as already mentioned.
This means
\begin{align}
N_\text{up}  < \frac{1}{6} \ln \frac{3}{\epsilon_\text{turn}}.
\end{align}
The second constraint is that the field range $\Delta \phi$ is limited for a given inflation model potential.
This condition is 
\begin{align}
N_\text{up}  \leq \frac{1}{3} \ln \frac{3 \Delta \phi}{\sqrt{2 \epsilon}}.
\end{align}
Combining them together, 
\begin{align}
N_\text{up}  \leq  \text{min} \left [  \frac{1}{3} \ln \frac{3 \Delta \phi}{\sqrt{2 \epsilon}} ,  \,  \frac{1}{6} \ln \frac{3}{\epsilon_\text{turn}} \right ].
\end{align}

Typically, $\Delta \phi  \simeq \sqrt{2 \epsilon} N$ up to an ${\cal O}(1)$ factor or logarithmic correction with respect to $N$ for a flat potential. 
For example, if the potential is a constant plus a small linear slope as in eq.~\eqref{V_expansion}, this is true. Also for the exponential expansion in eqs.~\eqref{V_exp-expansion} or \eqref{V_alpha-model}, it holds up to logarithmic corrections provided that the symmetry breaking term is not yet relevant at the turning point.
In such cases, the first term in the min bracket is given by $\frac{1}{3} \ln (3 N)$. Namely,
\begin{align}
N_\text{up}  \lesssim  \text{min} \left [  \frac{1}{3} \ln (3 N) ,  \,  \frac{1}{6} \ln \frac{3}{\epsilon_\text{turn}} \right ].
\end{align}
If we substitute $N= N_\text{CMB}$, the maximum e-folding during the rolling up stage is of order unity, $N_\text{up} \lesssim \mathcal{O}(1)$.   We conclude that it is unlikely that a large number of e-folding number is realized during the rolling-up motion.

\section{Shift symmetry breaking by a higher-order term} \label{sec:high-order-breaking}
In the main text, we consider the lowest possible order breaking term $e^{\lambda \phi}$.  It is possible that this term is absent or coefficients are negligible, and the dominant effects of shift symmetry breaking are due to a higher-order term $e^{n \lambda \phi}$ where $n$ is an integer.  In fact, such a higher-order term can have less impact on the CMB scale but it can rise more quickly to shorten the field range where the slow roll is possible.

In this Appendix, we consider the potential  $V= V_0 (1 - e^{-\lambda \phi} + \varepsilon_{\text{br}} e^{n\lambda \phi} )^2$, which sets the global property (e.g.~no local minima).  For our purposes for inflationary analyses, it can safely be approximated as
\begin{align}
V = V_0 \left( 1 - 2 e^{-\lambda \phi} + 2 \varepsilon_{\text{br}} e^{n\lambda \phi} \right),
\end{align}
where again $\lambda = \sqrt{2/3\alpha} \gg 1$.  Namely, we neglect higher-order terms in $e^{-\lambda \phi}$ and $\varepsilon_{\text{br}} e^{n \lambda \phi}$.

The slow roll parameters are 
\begin{align}
\epsilon = & 2 \lambda^2 \left( e^{-\lambda \phi} + n \varepsilon_{\text{br}} e^{n\lambda \phi} \right)^2, \\ 
\eta = & 2 \lambda^2 \left( - e^{-\lambda \phi} + n^2 \varepsilon_{\text{br}} e^{n\lambda \phi} \right), \\ 
\xi^2 = & 4 \lambda^4 \left( e^{-\lambda \phi} + n \varepsilon_{\text{br}} e^{n\lambda \phi} \right)  \left( e^{-\lambda \phi} + n^3 \varepsilon_{\text{br}} e^{n\lambda \phi} \right) .
\end{align}

The e-folding number $N$ is calculated as 
\begin{align}
N 
 =& \frac{e^{\lambda \phi}}{2\lambda^2} \, {}_2 F_1 \left [  1, \frac{1}{n+1}; \frac{n+2}{n+1}; - \varepsilon_{\text{br}} n e^{(n+1)\lambda \phi} \right ] -  {}_2 F_1 \left [  1, \frac{1}{n+1}; \frac{n+2}{n+1}; - \frac{c}{N_\text{CMB}^{n+1}}\right ],  \label{N_formula_n}
\end{align}
where ${}_2 F_1 (a,b;c;z)$ is the hypergeometric function, and we defined 
\begin{align}
c \equiv n \varepsilon_{\text{br}} \lambda^{2(n+1)} (2 N_\text{CMB})^{n+1},
\end{align}
as a parameter combination controlling the magnitude of the correction by symmetry breaking terms.
In eq.~\eqref{N_formula_n}, it is assumed that inflation ends quickly after the slow-roll ends, $|\eta | = 1$.  
If we set $n=1$, eq.~\eqref{N_formula_n} reduces to eq.~\eqref{N_formula_1}. Also, ${}_2F_1$ becomes unity in the limit $\varepsilon_{\text{br}} \to 0$.
Similarly, the e-folding number along the whole available slow-roll region $(\epsilon, \, |\eta| \leq 1)$ is given by
\begin{align}
N_\text{total} = & N_\text{CMB} \left( \frac{N_\text{CMB}}{n c} \right)^\frac{1}{n}  {}_2 F_1 \left [  1, \frac{1}{n+1}; \frac{n+2}{n+1}; - c  \left( \frac{N_\text{CMB}}{n c} \right)^{\frac{n+1}{n}} \right ]  \nonumber \\
&  \qquad \qquad \qquad -  {}_2 F_1 \left [  1, \frac{1}{n+1}; \frac{n+2}{n+1}; - \frac{c}{N_\text{CMB}^{n+1}}\right ].
\end{align}
Thus, the total e-folding number along the whole available slow-roll region is written in terms of the e-folding number at the CMB scale and the correction parameter $c$, which is determined once $\varepsilon_{\text{br}}$ and $\lambda$ (equivalently $\alpha$) are fixed.

We invert the relation between $N$ and $\phi$ to evaluate the inflationary observables at the CMB scale.
To this end, we recursively solve it by the perturbation method.
The results are
\begin{align}
\lambda^2 e^{-\lambda \phi_\text{CMB}} =& \frac{1}{2N_\text{CMB}} \left( 1 - \frac{c}{n+2} - \frac{(n^2+n-1) c^2}{(n+2)^2 (2n+3)} + \mathcal{O}(c^3)\right)  , \\ 
\lambda^2 \varepsilon_{\text{br}} e^{n\lambda \phi_\text{CMB}} =& \frac{1}{2 N_\text{CMB}} \left( c + \frac{n c^2}{n+2}  + \mathcal{O}(c^3)   \right) .
\end{align}
One can use these formulas to evaluate quantities such as $\ns$ and $\alpha_\text{s}$.

\begin{figure}[htb]
 \centering
   \subcaptionbox{$n = 2$ \label{sfig:n=2}}
{\includegraphics[width=0.49 \columnwidth]{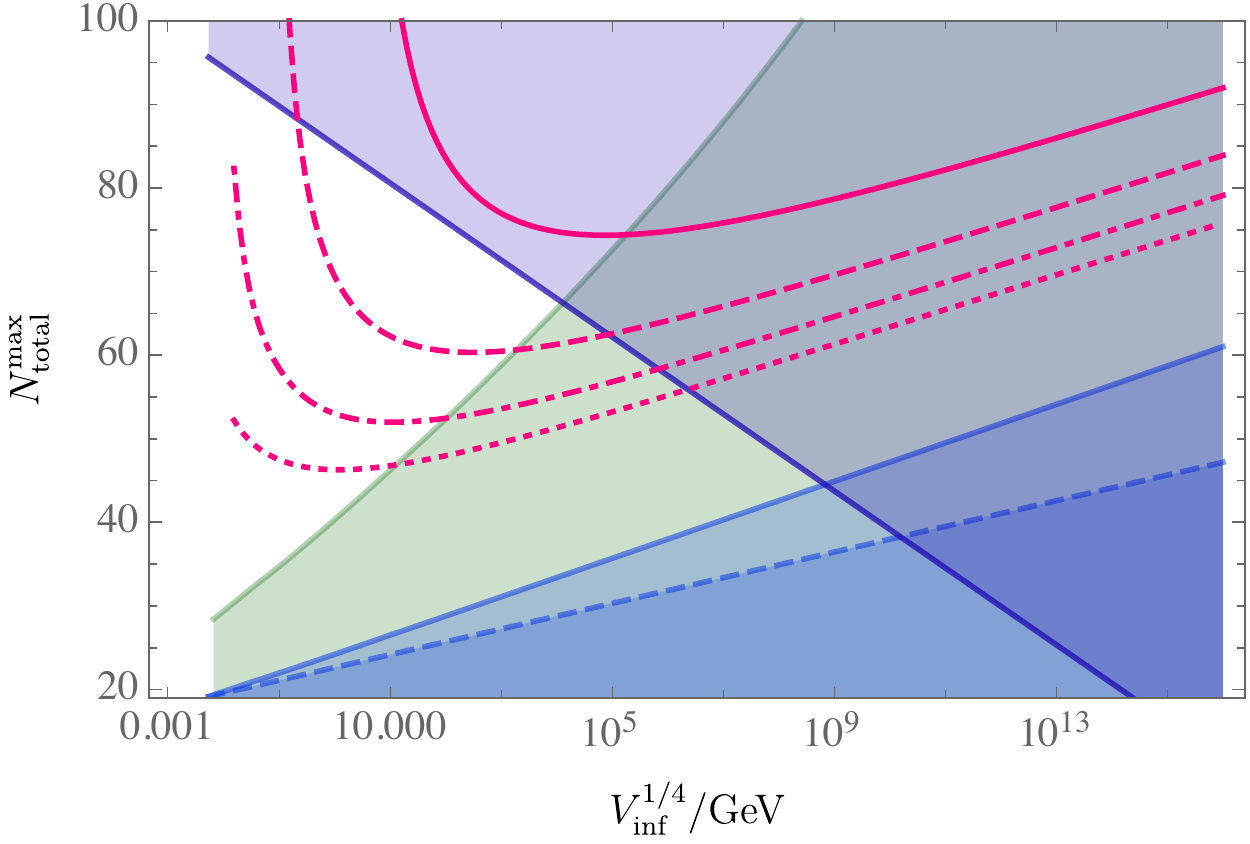}}~
   \subcaptionbox{$n = 3$ \label{sfig:n=3}}
{\includegraphics[width=0.49 \columnwidth]{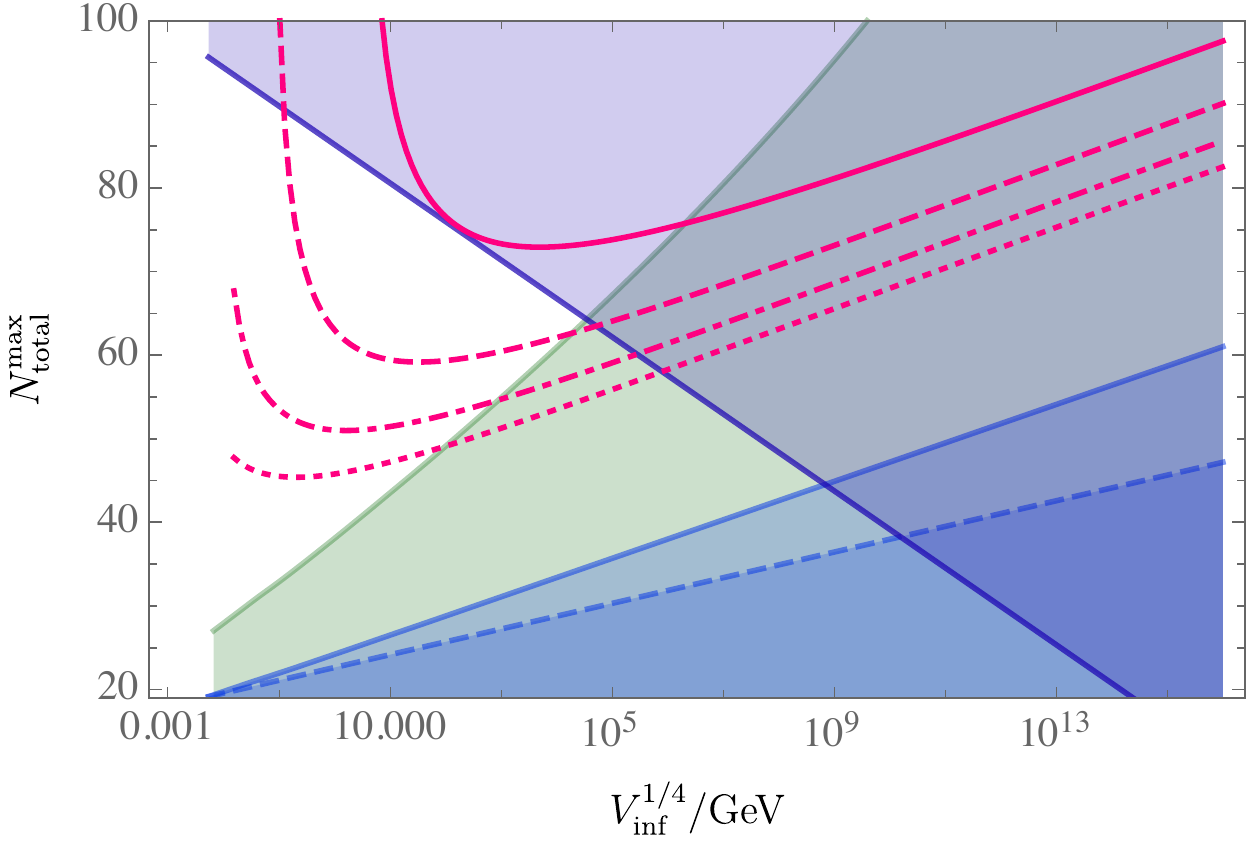}}~
  \caption{Same as Fig.~\ref{fig:N_range_n=1}, but for $n=2$ and 3, where $n$ is the power of the shift symmetry breaking term $\varepsilon_{\text{br}} e^{n \lambda \phi}$.}
 \label{fig:N_range_n}
  \end{figure}

The constraints on the allowed range of $N_\text{total}^{\text{max}}$ are shown in Fig.~\ref{fig:N_range_n}. From the intersection of the TCC constraint (purple line) and the additional possible e-folding number (green line) which respects $\ns$, we obtain an upper bound on the inflation energy scale and the tensor-to-scalar ratio
\begin{align}
V_\text{inf}^{1/4} <& \begin{cases}
7.0 \times 10^3 \, \text{GeV}  &  (n=1) \\
1.3 \times 10^4 \, \text{GeV}  &  (n=2) \\
3.5 \times 10^4 \, \text{GeV}  &  (n=3) 
\end{cases}
 , 
&
r < & \begin{cases}
2.2 \times 10^{-51}   &  (n=1) \\
2.6 \times 10^{-50}   &  (n=2) \\
1.4 \times 10^{-48}   &  (n=3) 
\end{cases} 
.
\end{align}.

\section{The maximum e-folding for thermal inflation} \label{sec:thermal}
In this independent Appendix, we discuss the
possible e-folding number, $N_{\rm TI}$,  in which inflation is only driven by interactions between the scalar field and the background plasma but not by the scalar potential at zero temperature (thermal inflation~\cite{Lyth:1995ka}). 

If the interactions are fast enough, the Universe can be described by 
the thermal field theory with a temperature $T$. 
For the length scale $\ell > 1/T$, the finite temperature effective potential of $\phi$, $V_T(\phi)$, is good enough to describe the evolution of $\phi$. 
As an example,  
$V_T(\phi)$  can be expanded as  
\begin{align} \label{eq:TI_pot}
V_T(\phi) = V_0 + \frac{1}{2} (\lambda T^2 - m^2)  \phi^2 + \cdots,
\end{align}
around the origin. Here $\lambda>0$ denotes the contribution
of the couplings between $\phi$ and the background plasma, while  
$-m^2 <0$, so that the origin
 is unstable at zero temperature. 
At high temperatures, $\phi$ is trapped at the origin by thermal mass terms, so it provides 
a constant energy density, $V_0$. 
Assuming that the thermal inflation ($\rho_{\rm tot}\simeq V_0$)  starts at $T=T_i$ ($\rho_r \lesssim V_0$) and finishes at $T=T_f$, 
the total e-folding number is given by 
\begin{align}
N_{\rm TI} = \ln\frac{a_f}{a_i}= \ln\frac{T_i}{T_f}.
\end{align}

Before evaluating it from the model, 
one can provide an upper bound on $N_{\rm TI}$ by the following argument. 
As the inflation continues, 
the mean free path between collisions, $\ell_\text{fr} \sim 1/T$,  in the thermal  plasma is exponentially increasing as
$\ell_\text{fr}(t) = e^{H t}/T_i$. 
If $\ell_\text{fr}(t)$ can be larger than the Hubble horizon, $1/H$, 
the assumption of thermal equilibrium is no longer valid.
Because the relaxation time becomes longer than the age of the Universe,
the plasma does not give a meaningful effect for the evolution of $\phi$.
Instead of using eq.~(\ref{eq:TI_pot}), we 
should calculate the evolution of $\phi$ mostly by the zero temperature scalar potential. Therefore the thermal inflation must end before $\ell_\text{fr}(t) =1/H$, which leads to the bound on $N_{\rm TI}$ as\footnote{
Thermal equilibrium may be maintained for a timescale larger than $1/H$ because of Gibbons-Hawking radiation $T\sim H/2\pi$~\cite{Gibbons:1977mu}.  However, the thermal mass is too small to stabilize $\phi$ at the origin, and it will be destabilized within a Hubble time mainly due to the zero temperature potential.  Thus, our conclusion is not changed.
} 
\begin{align} \label{TI_bound}
N_{\rm TI} < \ln\frac{T_i}{H}.
\end{align} 
In the present example, the curvature of the potential changes its sign 
at $T_f \simeq m$. Our starting point of this Appendix is that 
the inflation is only driven by thermal interactions, i.e.~$m\gg H
\sim V_0^{1/2}/M_{\rm P}$, so the bound \eqref{TI_bound} is satisfied.
Note that this inequality is confirmed without substituting the value of $T_i$, which is given by $T_i \simeq V_0^{1/4}$ in thermal inflation.

Here, we can find an interesting analogy between the arguments for inequalities  (\ref{TCC}) and (\ref{TI_bound}).
As $1/M_{\rm P}$ is the minimum length scale for which the quantum field theory (QFT) with weakly coupled gravity is valid,
$1/T_i$ is also the minimum length scale for which the thermal dynamic  treatment is valid when the inflation starts. 
In both cases, 
$1/H$ plays the role of the causal horizon. 
A difference is that when inflation is driven by thermal interactions, 
we know the microscopic theory for  $\ell < 1/T$: Non-equilibrium QFT, so we can calculate the evolution of the scalar field 
and understand why thermal inflation should end before $N_{\rm TI} = \ln T_i/H$. 
On the other hand, for the TCC case, we do not know the correct underlying theory for $\ell < 1/M_{\rm P}$, which means that we cannot figure out the true origin of the inflation and why the inflation should end before $N = \ln M_{\rm P}/H$. The TCC is just a conjecture at this moment.

\small

\bibliographystyle{utphys}
\bibliography{tcc.bib}

\end{document}